\documentclass[pra,twocolumn,showpacs,superscriptaddress,
amsmath,amssymb]{revtex4}

\usepackage{graphicx}
\usepackage{dcolumn}
\usepackage{bm}
\usepackage{epsfig}
\usepackage{amsmath}
\usepackage{color}

\newcommand{\gam}           {\gamma}
\newcommand{\kap}           {\kappa}
\newcommand{\wa}    {\Delta_{a}}
\newcommand{\wc}    {\Delta_{c}}
\newcommand{\wat}    {\widetilde{{\omega}}_{a}}
\newcommand{\wct}    {\widetilde{{\omega}}_{c}}

\newcommand{\Lsig}  {L_{\sigma}}

\newcommand{\La}  {L_{a}}

\newcommand{\aop}        {a}
\newcommand{\maop}        {\left<\aop\right>}

\newcommand{\caop}       {a^{\dag}}

\newcommand{\csig}       {\sigma^{\dag}}
\newcommand{\sig}       {\sigma}
\newcommand{\msig}        {\left<\sig\right>}
\newcommand{\mcsig}        {\left<\sig\right>^*}

\newcommand{\vecr}       {\bm{r}}
\newcommand{\veck}       {\bm{k}}
\newcommand{\DD}       {{\textrm D}}
\newcommand{\MM}       {{\textrm M}}
\newcommand{\SSS}       {{\textrm S}}
\newcommand{\ee}       {{\textrm e}}

\newcommand{\dd}   {\hbox{\textrm d}}

\newcommand{\beq}   {\begin{equation}}
\newcommand{\eeq}   {\end{equation}}
\newcommand{\bseq}   {\begin{subequations}}
\newcommand{\eseq}   {\end{subequations}}
\newcommand{\beqz}  {\setlength{\mathindent}{0cm}\begin{equation}}
\newcommand{\eeqz}  {\end{equation}}
\newcommand{\ber}   {\begin{eqnarray}}
\newcommand{\eer}   {\end{eqnarray}}
\newcommand{\bers}  {\begin{eqnarray*}}
\newcommand{\eers}  {\end{eqnarray*}}

\begin{document}

\title{Momentum diffusion for coupled atom-cavity oscillators}

\author{K. Murr}
\email{karim.murr@mpq.mpg.de}
\author{P. Maunz}
\author{P.W.H. Pinkse}
\author{T. Puppe}
\author{I. Schuster}
\affiliation{Max-Planck-Institut f\"ur Quantenoptik,
Hans-Kopfermann-Strasse~1, D-85748 Garching, Germany}
\author{D. Vitali}
\affiliation{Dipartimento di Fisica, Universit\`a di Camerino,
I-62032 Camerino, Italy}
\author{G. Rempe}
\email{gerhard.rempe@mpq.mpg.de}
\affiliation{Max-Planck-Institut f\"ur Quantenoptik,
Hans-Kopfermann-Strasse~1, D-85748 Garching, Germany}

\date{\today}

\begin{abstract}
It is shown that the momentum diffusion of free-space laser cooling has
a natural correspondence in optical cavities when the internal state of
the atom is treated as a harmonic oscillator. We derive a general
expression for the momentum diffusion which is valid for most
configurations of interest: The atom or the cavity or both can be probed
by lasers, with or without the presence of traps inducing local atomic
frequency shifts. It is shown that, albeit the (possibly strong)
coupling between atom and cavity, it is sufficient for deriving the
momentum diffusion to consider that the atom
couples to a mean cavity field, which gives a first contribution, and
that the cavity mode couples to a mean atomic dipole, giving a second
contribution. Both contributions have an intuitive form and present a
clear symmetry. The total diffusion is the sum of these two
contributions plus the diffusion originating from the fluctuations of
the forces due to the coupling to the vacuum modes other than the cavity
mode (the so called spontaneous emission term).
Examples are given that help to evaluate the heating rates induced by
an optical cavity for experiments operating at low atomic saturation.
We also point out intriguing situations where the atom is heated
although it cannot scatter light.
\end{abstract}

\pacs{42.50.Lc, 32.80.Lg, 42.50.Ct, 32.80.-t}

\maketitle
\section{Introduction}
\begin{figure}
\includegraphics[width=4cm]{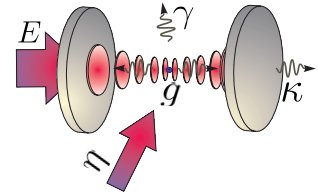}
\caption{\label{Setup} (Color online) An atom placed between two
mirrors is excited from the side (laser-atom Rabi frequency $2\eta$).
The cavity mode is also coherently pumped by a laser source
(coupling strength $E$).
The mirrors dissipate the field at a rate $\kap$,
which is the analogue of the relaxation rate $\gam$ of the atom's
dipole moment. The atom-cavity vacuum Rabi frequency is denoted by $2g$.
Forces exist mainly along the cavity axis
(due to $E$ and/or $\eta$), and along the side-laser beam.}
\end{figure}

A single atom coupled to a single mode of an optical cavity is the
archetype model of dissipative electrodynamics.
This system provides a matter-light interface for fundamental studies as
well as applications.
It has two well-known characteristics (see Fig.~\ref{Setup}).
First, the cavity mirrors confine the light, which can lead to a strong
influence of the atom on the cavity field and vice versa.
Second, in addition to the atom emitting photons into free space,
the cavity mirrors now offer an extra loss channel.
The spatially directed light can easily be detected to observe the
properties of the system. The new dissipative channel is also
responsible for a new dissipative force, which can efficiently damp
atomic motion
\cite{Mossberg91,Zaugg94,Doherty97,Horak97,Hechenblaikner98,Vuletic00,
Enk01,Vuletic01,Domokos01,Fischer01,Domokos02,Murr03,Domokos04,
Maunz04,Nussmann05}. However, dissipation inevitably leads to
fluctuations, in particular light-force fluctuations, which heat the
atom.

Indeed, several experiments in the regime of strong coupling and low
atomic excitation reported excessively large heating rates
\cite{muenstermann99,Ye99,McKeever03,Boca04,Maunz05}, up to two orders
of magnitude larger than for an atom in free-space laser fields. The
physical origin of this is obscured: Expressions that were
calculated for the heating rates are restricted to specific
geometries. They are not accessible to an intuitive interpretation as
they mix those two effects of confinement and dissipation, and contain
correction terms to the heating that could be negative.

This paper derives a general expression for the heating rate which also
gives insight into the physics of force fluctuations in a cavity: By
effectively decoupling the atom-cavity system, we first show that
the equation for the momentum diffusion can be written as the sum of
positive contributions, where we clearly identify the
fluctuations of the atomic dipole and the fluctuations of the cavity
field. Second, this form is intuitive as it allows to perceive the
heating in free space \cite{Cook80,Cohen92} and in confined space
alike. While our expression generalizes previous ones
\cite{Horak97,Hechenblaikner98,Domokos01,Fischer01,Enk01,Vuletic01,
Domokos02}, a powerful new feature is its invariance: It applies,
regardless of whether the atom and/or the cavity is probed,
regardless of the spatial structure of the involved light fields,
with or without the inclusion of Stark shifts due to an additional
dipole-trap, and could be naturally extended to an arbitrary number of
atoms and modes, linear or ring cavities.
We explain that such an invariance is a signature of harmonic
oscillators, a description which
applies to the experimentally most relevant regime of vanishing atomic
excitation. Lastly, we point out some conceptually interesting
situations, like the persistence of heating induced by cavity-field
fluctuations even if the cavity field is in the vacuum state
$|0\rangle$.

\section{Free-space momentum diffusion at low saturation}
From the theory of Brownian motion, the random character of a
force leads to a spread of the particle's momentum $\bm{p}$.
This heating mechanism is characterized by the (momentum)
diffusion coefficient $2\DD=\dd/\dd t\langle(\Delta\bm{p})^2\rangle$.
For a particle of mass $m$ the heating rate is then given by $\DD/m$
(energy per unit time).
For a two-state atom at rest and coupled to a single-mode laser, the
diffusion in the harmonic description, which requires a low excited
state $|e\rangle$ occupation probability $P_e$, reads
\cite{Gordon80,Cohen92}
\beq
\label{DatTotFree}
2\DD= (\hslash k)^22\gam P_e + |\hslash\nabla\msig|^22\gam\ .
\eeq
The first term is the familiar contribution arising from the random
direction of the spontaneously emitted photons (with momentum
norm $\hslash k$), occurring at a rate $2\gam P_e$, where
$1/2\gam$ is the lifetime of the upper state of the atom. The second
term stems from the fluctuations of the laser force, and is
determined by the gradient of the atomic mean coherence $\msig$
(defining the dipole moment). At low atomic saturation
$P_e\approx|\msig|^2\ll1$, the
mean coherence is given by $\msig=\eta(\vecr)/(\wa-i\gam)$, where
$2|\eta(\vecr)|$ is the laser-atom Rabi frequency at the position
$\vecr$ of the center of mass of the atom and $\wa=\omega_{eg}-\omega_L$
is the detuning between the transition frequency $\omega_{eg}$ of the
atom and the laser frequency $\omega_L$.

For a laser running wave, $\eta(\vecr)=\eta_0\ee^{i\veck_L\cdot\vecr}$,
representing a stream of photons all propagating with the same momentum
$\hslash\veck_L$ ($\omega_L=c|\veck_L|$), that diffusion originates from
the random character of the absorption process \cite{Cook80,Cohen92}.
Per unit time, the atom absorbs a random number of photons, and hence
acquires a random momentum, along the laser beam. Despite all laser
photons having the same momentum, the atom thus heats, as long as
there is light. For a laser standing wave,
$\eta(\vecr)=\eta_0\cos(\veck_L\cdot\vecr)$, the momentum diffusion
can be simply understood in terms of absorption and emission of
photons only when the diffusion is averaged over a spatial
period. In this case the standing wave is seen as two independent
running waves of half intensity, and the diffusion is just the sum
of each contribution \cite{Cohen92}. Otherwise, interpreting the
momentum diffusion in terms of ``recoil kicks'' for any location in
a standing wave is not trivial. A known intriguing example which
is discussed in detail in \cite{Cohen92} (also see \cite{Gordon80})
is that of an atom at a node. At a node, there is no light,
$\eta(\vecr)=0$, meaning that the atom does not emit photons,
$P_e=0$, but the diffusion is finite $\DD\neq0$.

Besides, we recall that the diffusion coefficient derived within the
dressed-state picture \cite{Dalibard85} is
not the one discussed here Eq.~(\ref{DatTotFree}).
More precisely, the term derived from the harmonic description
$|\hslash\nabla\msig|^22\gam$, and the term obtained from the
dressed-state picture \cite{Dalibard85} are distinct limits of the full
expression for the diffusion \cite{Gordon80}.
Quantitatively, Eq.~(\ref{DatTotFree}) dominates the dressed-state one
for $P_e\ll1$ and $2P_e\ll(\gam/\wa)^{2/3}$ \cite{Gordon80}.
Those inequalities also show that for large detuning $\wa$ and
sufficiently high intensity the dressed-state term \cite{Dalibard85} can
be dominant at low atomic saturation. For example, for $\wa=10^3\gam$
and $P_e=0.01$, one has $P_e\ll1$ but $2P_e>(\gam/\wa)^{2/3}$. However,
for most situations of interest the limit of low saturation coincides
with the harmonic limit considered here.

\section{Momentum diffusion in the presence of a Cavity}
We show that in a cavity, and independently from which way the
system is excited, the diffusion for coupled atom-cavity oscillators
can be written in the following invariant form:
\beq
\label{DatTot}
2\DD=(\hslash k)^22\gam P_e+|\hslash\nabla\msig|^22\gam +
|\hslash\nabla\maop|^22\kap\ .
\eeq
The mean coherence $\msig$ ($P_e\!\!\approx\!\!|\msig|^2$) is now
evaluated within the cavity setting. The additional term
$|\hslash\nabla\maop|^22\kap$ comes from the fluctuations of the cavity
mode. It depends only on the
gradient of the steady-state amplitude of the cavity field $\maop$,
and is proportional to the decay rate $2\kap$ at which photons
escape through the mirrors. The momentum diffusion is completely
determined from the expectation values $\msig$ and $\maop$,
which depend on the position of the atom and which contain all the
information on the coupling between atom and cavity, see
Eqs.~(\ref{MsigMaop}), below.

Let us now derive Eq.~(\ref{DatTot}) and analyze and interpret the
relative contributions of the different terms. The derivation of
Eq.~(\ref{DatTot}) is obtained via two unconventional
ways. The first way is to start from the well known driven
Jaynes-Cummings Hamiltonian and to perform a particular
mathematical operation which corresponds to transferring the information
on the dynamics into the components of a new force operator.
The second way shows that Eq.~(\ref{DatTot}) is also the momentum
diffusion obtained from two decoupled Hamiltonians, one describing the
interaction of an atom coupled to a mean cavity field, and the other one
describing the interaction of a cavity mode coupled to a mean atomic
dipole \cite{Murr03}. The second way gives us a natural interpretation
and also shows that one can truncate the Jaynes-Cummings Hamiltonian
without modifying the result for the diffusion coefficient.

\subsection{Derivation of Eq.~(\ref{DatTot}) as a solution from
the extended Jaynes-Cummings Hamiltonian}
One first starts from a general driven atom-cavity system, described by
the extended Jaynes-Cummings Hamiltonian (labeled ``$JC$'')
\ber
\label{HJC}
H_{JC}/\hslash=&&\wa\csig\sig
-\eta(\vecr)\csig
-\eta^*(\vecr)\sig\nonumber\\
&+&\wc\caop\aop + E\caop+E^*\aop\\
&+&g(\vecr)\aop\csig+g^*(\vecr)\caop\sig\nonumber\ ,
\eer
where the atom, with lowering operator $\sig$, is dipole-coupled
(last line) to a single cavity mode of creation operator $\caop$.
Here $2g$ is the atom-cavity mode vacuum Rabi frequency.
We omit writing the kinetic term $\bm p^2/2m$ because the position
$\vecr$ is (ultimately) treated classically and we focus on the
lowest order in the velocity $\bm v$, $k_Lv\ll(\gam,\kap)$
(atom almost at rest).
Atom and mode are coherently excited by two different near-resonant
classical laser sources, that have the same frequency $\omega_L$, $E$ is
the coupling strength of the axial laser and we recall that $2|\eta|$ is
the side laser-atom Rabi frequency. The strength $E$ is defined such
that $|E|^2/(\wc^2+\kap^2)$ is the empty cavity photon number, where
$\wc=\omega_{cav}-\omega_L$ is the detuning between a cavity resonance
frequency $\omega_{cav}$ and the laser(s).
We assumed the rotating-wave approximation
and wrote the Hamiltonian in the interaction picture with
respect to the laser frequency $\omega_L$.
The evolution of the reduced density matrix $\rho(t)$ for the
atom-cavity system must account for the loss mechanisms
(atom and cavity decay):
$\dot\rho=-i[H_{JC},\rho]/\hslash+\kap\La\rho+\gam\Lsig\rho$, where
$\La\rho=2\aop\rho\caop-\rho\caop\aop-\caop\aop\rho$, and for the atom
$\Lsig\rho=2\int\!\!d^2\hat{\bm{k}}N(\hat{\bm k})
\ee^{-i\bm k\cdot\vecr}\sig\rho\csig
\ee^{+i\bm k\cdot\vecr}-\rho\csig\sig-\csig\sig\rho$, where
$\hat{\bm{k}}=\veck/k$ ($k=\omega_{eg}/c$) is the direction of
spontaneously emitted photons, with the angular distribution
$N(\hat{\bm{k}})$ (dipole pattern),
$\int\!\!d^2\hat{\bm{k}}N(\hat{\bm k})=1$.

We now show that Eq.~(\ref{DatTot}) is the correct diffusion
obtained from (\ref{HJC}) if the internal state of the atom is treated
as a harmonic oscillator $[\sig,\csig]\rightarrow\openone$
(no saturation effects). The conventional procedure
\cite{Gordon80,Cohen92,Hechenblaikner98,Fischer01,Domokos02}
leads to a first expression, which we write in a compact form
$2\DD\!=\!(\hslash k)^22\gam P_e
-\bm u^\dag(\MM^{-1}\!\!+\!\MM^{-1\dag})\bm u$.
The $2\!\times\!2$ matrix $\MM\!=\!\textrm P\!-\!i\textrm G$ determines
the dynamics of coupled oscillators, where
$\textrm P\neq\textrm P^\dag$ is
the diagonal matrix for decoupled oscillators
$(\textrm P_{11}=-i(\wa-i\gam),\textrm P_{22}=-i(\wc-i\kap))$,
and $\textrm G=\textrm G^\dag$ is the coherent coupling off-diagonal
Hermitian matrix ($\textrm G_{12}=g$). Already in limiting cases,
this expression is complicated
\cite{Horak97,Hechenblaikner98,Fischer01,Domokos02}.
Our first clarification is based on the
following observation: The vector $\bm u^\dag$ is the coordinate
of the force fluctuation
$\delta\bm F_{JC}=\bm u^\dag{\textrm w}+\textrm w^\dag\bm u$
$+[\mathcal{O}]$ in the subspace
$\textrm w=(\delta\sig,\delta\aop)^T$ of the fluctuations
$\delta\sig=\sig-\msig$ and $\delta\aop=\aop-\maop$. The
($\bm u^\dag,\bm u$) are the only coordinates of $\delta\bm F_{JC}$ that
contribute to the diffusion in the harmonic limit.
Writing $\bm F_{JC}=-\nabla H_{JC}$, and then expressing ($\sig,\aop$)
in terms of $(\delta\sig,\delta\aop)$, one obtains
$\bm u=\hslash[\nabla\textrm I-i(\nabla\MM)\SSS]$, where
$\textrm I=(\eta,-E)^T$ and $\SSS=(\msig,\maop)^T$. Now, in steady
state one has $\textrm I-i\MM\SSS=0$, which by differentiation gives
$\bm u=i\hslash\MM\nabla\SSS$: Hence
$2\DD=(\hslash k)^22\gam P_e-\hslash\nabla\SSS^\dag(\textrm P\!\!\!$
$+\textrm P^\dag)\hslash\nabla{\SSS}$. The matrix
$\textrm P^\dag+\textrm P$ is diagonal, with $-2\gam$ and $-2\kap$ as
elements. This is Eq.~(\ref{DatTot}). The diffusion tensor in three
dimensions exhibits a similar structure as Eq.~(\ref{DatTot}) provided
it is defined symmetric with respect to the spatial coordinates.
A closer inspection also shows that Eq.~(\ref{DatTot}) still holds if
any of the frequencies ($\wa,\eta,\wc,E,g$) depends on the atomic
position. In particular, one can include the presence of a dipole trap
through local Stark shifts $\wa\rightarrow\wa(\vecr)$.

\subsection{Derivation of Eq.~(\ref{DatTot}) as a solution from
decoupled Hamiltonians}

Equation~(\ref{DatTot}) can be derived in a more direct way by
establishing a correspondence to the ``mean-field'' Hamiltonian
$H=H_{atom}\!+\!H_{mode}\!+\!U$ \cite{Murr03},
\bseq
\label{Hatom-mode}
\ber
\label{Hatom}
H_{atom}/\hslash=&&\wa\csig\sig
- \left(\,[\eta - g\maop]\csig +h.c.\right)\ ,\\
\label{Hmode}
H_{mode}/\hslash=&&\wc\caop\aop
+ \left(\,[E + g^*\msig]\caop +h.c.\right)\ ,
\eer
\eseq
with the mean interaction energy $U=-\hslash g\maop\mcsig +h.c.$.
Note that, since the expectation values are dynamical variables,
derivation of the friction force requires $U$.
However, for our purposes the atom is
(almost) at rest, i.e. the mean values $\maop$ and $\msig$ are fixed
and the Hamiltonian $H$ (\ref{Hatom-mode}) decouples. Our
Hamiltonian and the Jaynes-Cummings one lead to the same expression
for the momentum diffusion, Eq.~(\ref{DatTot}). Indeed, provided
that one defines the force as $\bm F=-\nabla H$, equation
(\ref{DatTot}) is again obtained, but now straightforwardly:
Eq.~(\ref{Hatom}) gives
\bseq
\beq
2\DD_{atom}=|\hslash\nabla\msig|^22\gam\ , \eeq and
Eq.~(\ref{Hmode})
\beq 2\DD_{mode}=|\hslash\nabla\maop|^22\kap\ ,
\eeq
\eseq
and therefore Eq.~(\ref{DatTot}),
$2\DD=(\hslash k)^22\gam P_e+2\DD_{atom}+2\DD_{mode}$.

The interpretation of Eq.~(\ref{DatTot}) follows immediately from that
of the mean field Hamiltonian Eqs.~(\ref{Hatom-mode}).
In the first Hamiltonian (\ref{Hatom}), the atom ``sees'' a total field
which is the incident source field
$\eta$ and a mean cavity field $\maop$. Thus it experiences
an effective coupling $\Omega_{atom}(\vecr) = \eta(\vecr) -
g\left<a\right>(\vecr)$, similar to free space. Symmetrically, the
cavity field
``sees'' a source field of strength $E$ and a mean atomic
dipole $\msig$. Thus, the cavity mode is driven with an
effective strength
$\Omega_{mode}(\vecr)=  E + g^*\!\!\left<\sig\right>(\vecr)$.
One thus obtains a general picture where each object is coupled to a
classical-averaged partner, and fluctuates.
We discuss this picture in a specific example in
Sec.~\ref{Interpretations}.

The simple form of Eq.~(\ref{DatTot}) may suggest that the result
could have been guessed from the beginning, the argument being that
the driven Jaynes-Cumming Hamiltonian for an atomic oscillator is
already symmetric if one exchanges atom and cavity variables so that
every physical quantity must reflect this symmetry
$(\wa,\gam,-\eta,g,\sig)\leftrightarrow(\wc,\kap,E,g^*,\aop)$. This
argument is not sufficient because terms of the form
$\propto\Re[(\hslash\nabla\mcsig)(\hslash\nabla\maop)]$
or $\propto(\wc\gam+\wa\kap)$ do all satisfy the symmetry of coupled
oscillators, but do not appear in Eq.~(\ref{DatTot}).
It is already not obvious that the diffusion coefficient is a function
of only $\msig$ and $\maop$ (and $\kap,\gam$).
Also, cross terms of the form
$\propto\Re[(\hslash\nabla\mcsig)(\hslash\nabla\maop)]$ could have been
expected because of the form of the dipole-coupling Hamiltonian,
$g\aop\csig+h.c.$.
It is possible to show that the form of Eq.~(\ref{DatTot})
follows from a general property of integrals of time-symmetric
autocorrelation functions for dipole-coupled oscillators
(which is not a property satisfied by all observables).
This property will not be detailed here, rather we
express it through the following subtle point:
The original Hamiltonian (\ref{HJC}) can be expressed as the
second one (\ref{Hatom-mode}), plus a term,
$H_{JC}=H+(g\delta\aop\delta\csig+h.c.)$. \emph{The additional term
$g\delta\aop\delta\csig+h.c.$ has an essential role in the dynamics
as it gives large contributions to the diffusion derived from the
couple ($H_{JC},\bm F_{JC}$). However, this additional term can be
omitted, yielding $H$, but one must then redefine the force operator as
$\bm F=-\nabla H$ such that the couple ($H,\bm F$) still gives
Eq.~(\ref{DatTot}).}
In other words, Eq.~(\ref{DatTot}) can be obtained if one truncates
(\ref{HJC}), which means that atom and mode are dynamically decoupled,
but the force must be modified too; it now depends on the position
$\vecr$ also through $\msig$ and $\maop$.
This is the main reason why the identification of our result with
respect to existing expressions
\cite{Horak97,Hechenblaikner98,Domokos01,Fischer01,Domokos02,Murr03}
is not immediate.
\section{Evaluating the diffusion}
The momentum diffusion is completely determined given the
expectation values $\msig=\Omega_{atom}/\wat$ (with
$\wat\!=\!\wa\!-i\gam$), and $\maop=-\Omega_{mode}/\wct$ (with
$\wct\!=\!\wc\!-\!i\kap$), which gives:
\bseq
\label{MsigMaop}
\ber
\label{Msig}
\msig&=&\frac{1}{1-\nu(\vecr)}
\big[\ \,\eta(\vecr)+g(\vecr)\frac{E}{\wct}\ \,\big]/\wat\ ,\\
\label{Maop}
\maop&=&-\frac{1}{1-\nu(\vecr)} \big[\
\,E+g^*(\vecr)\frac{\eta(\vecr)}{\wat}\ \,\big]/\wct\ ,
\eer
\eseq
where $\nu(\vecr)$ is the generalized cooperativity parameter
$\nu=|g|^2/(\wat\wct)$ \cite{Murr03}.
For coupled atom-cavity oscillators, those expectation
values are solutions to both Hamiltonians,  (\ref{HJC}) and
(\ref{Hatom-mode}) \cite{Murr03}. To obtain the momentum diffusion
one calculates the gradients $\nabla\msig$ and
$\nabla\maop$, the position dependence being in
$g(\vecr), \nu(\vecr), \eta(\vecr)$ and possibly $\wa(\vecr)$.

A parameter regime which is often considered theoretically as well
as experimentally is that of large atomic detuning $\wa$ and low
cavity detuning $\wc$. For large $|\wa|$ the atomic excitation is
suppressed so that diffusion due to photon emission into free space
can be neglected, whereas a low value for $|\wc|$ would enhance the
scattering into the cavity. While in this regime a strong friction
force exist, there is also a strong heating rate which we now
estimate. Let us first consider the situation where the cavity is
pumped ($E\neq0,\eta=0$) and we focus on a linear cavity
$g(\vecr)=g_0\cos(\veck_{cav}\cdot\vecr)$. Assuming $\wc=0$, and
$|\wa|\gg (|g|^2/\kap,\gam)$ (such that $|\nu|\rightarrow0$), one
obtains to lowest order in $1/\wa$, $\nabla\maop=(\nabla\nu)E/i\kap
+\mathcal{O}(1/\wa^2)$, and $\nabla\msig=-(\nabla g)E/i\kap\times
1/(\wa-i\gam) +\mathcal{O}(1/\wa^2)$. This gives
$\DD_{mode}\approx8C\DD_{atom}$, where $C=|g|^2/2\kap\gam$ is the
cooperativity parameter. We then notice that for
$|\wa|\gg (|g|^2/\kap,\gam)$, the value of the diffusion $\DD_{atom}$ is
close to that in free space for a Rabi frequency equal to $2gE/\kap$
(recall that $|E|^2/\kap^2$ is the number of photons in the cavity
without the atom). When spatially averaged along the cavity axis, the
spontaneous emission term is also equal to $\overline{\DD_{atom}}$.
One therefore sees that for an equivalent excitation probability the
averaged heating rate for an atom in a cavity is
$1+C_0$ times larger than the total free space
value, where $C_0=|g_0|^2/2\kap\gam$ is $C$ taken for maximum coupling
$g=g_0$ (antinode) \cite{Fischer01}. For current optical cavities this
factor can reach $C_0\approx10^2$. That the momentum diffusion
for an intracavity atom is $C$ times larger than the free space
value is a quite general rule in the regime of large $|\wa|$ and low
$|\wc|$, with numerical factors depending on how the system is
probed and on whether spatial averaging is invoked. For example, if
the atom is probed directly ($E=0,\eta\neq0$), the gradient that
dominates along the cavity axis is
$\nabla\maop=(\nabla g)/i\kap\times\eta/(\wa-i\gam)
+\mathcal{O}(1/\wa^2)$, while $\nabla\msig$ is at least of order
$\mathcal{O}(1/\wa^2)$. Therefore, the term
$\overline{\DD_{mode}}\approx C_0(\hslash k)^2\gam P_e$ is $C_0$ times
larger than the spontaneous emission term.

\section{\label{Interpretations}Interpretations and discussion}
\begin{figure}
\includegraphics[width=7.1cm,height=5cm]{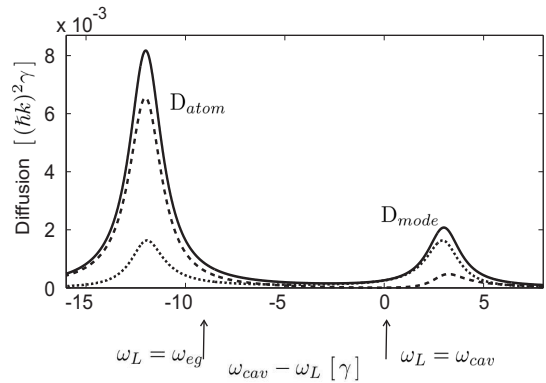}
\caption{\label{DNorm}Diffusion $\DD_{atom}+\DD_{mode}$ (solid)
along the laser axis when the atom is excited from the side,
for a running wave laser, as a function of the
laser frequency $\omega_L$,
($E=0, |\eta|=0.1\gam, \kap=\gam, g_0=6\gam,
\omega_{eg}-\omega_{cav}=9\gam$).
Diffusion due to spontaneous emission is equal to $\DD_{atom}$.
The peaks display the normal modes, $\DD_{atom}$ (dashed)
dominating for $\omega_L\approx\omega_{eg}$ and far from the
cavity resonance $\omega_{cav}$ (fluctuating atom), and vice versa
for $\DD_{mode}$ (dotted) dominating for
$\omega_L\approx\omega_{cav}$ (fluctuating mode).}
\end{figure}
Typically, the dynamics of
two coupled oscillators is dominated by the one which is closer to
resonance with the laser (e.g. $\DD_{mode}\gg\DD_{atom}$ for
$|\wc|\ll|\wa|$). This is illustrated in Fig.~\ref{DNorm} where the
normal-mode spectrum appears from the heating rate (at
the normal modes we have $\nu\rightarrow 1$, i.e.,
$\wa\wc\approx|g|^2$).
On the right peak, we have a fluctuating cavity field coupled to a
mean atomic dipole, while on the left peak the diffusion is
dominated by the fluctuations of an atomic dipole coupled to a
mean cavity field.
It is instructive to compare Fig.~\ref{DNorm} with Fig.~4 of
Ref.~\onlinecite{Maunz05}, where the normal-mode spectrum
(or vacuum Rabi splitting) appeared from the
measurement of (the inverse of) the time a trapped atom remained in the
cavity. The dramatic reduction of the storage time in this experiment
has been attributed to the large value of the diffusion
$\DD_{atom}+\DD_{mode}$ on the peaks. As we saw, those two terms have
relative contributions depending on the detunings and hence give insight
into the origin of the appearance of the peaks. Notice that in
\cite{Maunz05} the cavity is probed whereas the plot here is for atom
pumping. In this case the height of the two peaks in Fig.~\ref{DNorm}
should be interchanged and one would focus on the diffusion along the
cavity axis. This does not change the conclusions regarding the
distribution of $\DD_{atom}$ and $\DD_{mode}$ on their respective
peaks.

\begin{figure}
\includegraphics[width=7.1cm,height=5cm]{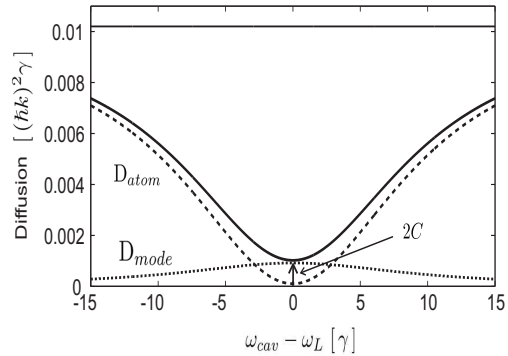}
\caption{\label{DSup}Diffusion $\DD_{atom}+\DD_{mode}$ (solid)
along the side laser, resonant to the atom
$\omega_L=\omega_{eg}$, versus the cavity frequency $\omega_{cav}$,
($E=0, |\eta|=0.1\gam, \kap=\gam, g_0=3\gam$).
The diffusion is well suppressed compared to free space for the same
intensity (horizontal line). $\DD_{mode}$ (dotted) reaches a maximum for
$\omega_{cav}=\omega_L$, where one has
$\DD_{mode}=2C\times\DD_{atom}\gg \DD_{atom}$ (dashed), and decreases
when the light hardly builds up inside the cavity
$\omega_{cav}\neq\omega_{L}$.}
\end{figure}

For a running wave laser in free space
$\eta(\vecr)=\eta_0\ee^{i\veck_L\cdot\vecr}$, the diffusion stems from
the random character of the absorption process \cite{Cook80,Cohen92}.
If $N_{abs}$ is the number of photons absorbed during a time
$t\gg 1/2\gam$, the atom gains the momentum
$\bm p=\hslash\veck_L N_{abs}$ (ignoring spontaneous and stimulated
emission). In steady state, $\langle N_{abs}\rangle=2\gam P_et$ is given
by the spontaneous emission rate $2\gam P_e$.
Since $N_{abs}$ is a random variable, the momentum spreads as
$\Delta\bm p^2=(\hslash\veck_L)^2\Delta N_{abs}^2$.
Now, in the harmonic description, the photon statistics follow the
Poisson law, $\langle\Delta N_{abs}^2\rangle=\langle N_{abs}\rangle$.
Hence the second term of Eq.~(\ref{DatTotFree})
$|\hslash\nabla\msig|^22\gam=(\hslash\veck_L)^22\gam P_e$.
For comparison, we look at the diffusion along the laser axis,
orthogonal to the cavity. Defining
$\bm F_{(a,m)}=-\langle\nabla H_{(atom,mode)}\rangle$
(here $U=const.$) gives $\bm F_{a}=\hslash\veck_L2\gam P_e$ and
$\bm F_{m}=\hslash\veck_L2\kap N_{cav}$, where
$N_{cav}=\langle\caop\aop\rangle$ is the cavity photon number.
Thus, the radiation pressure force acting on the atom
$\langle\bm F\rangle=\hslash\veck_L(2\gam P_e+2\kap N_{cav})$
is proportional to the rate at which photons are removed from the laser
$\dd\langle N_{abs}\rangle/\dd t=2\gam P_e+2\kap N_{cav}$: The diffusion
along the side laser is due to the statistics of photons that are
removed from the beam, in short, absorbed by the atom-cavity
system. That diffusion is plotted in Fig.~\ref{DNorm} and
Fig.~\ref{DSup}. As shown in Fig.~\ref{DSup}, for the same laser
intensity, the diffusion is suppressed with respect to free space.
This is due to the suppression of the atomic fluorescence
($|1-\nu|^2\gg 1$). Otherwise, as discussed previously, for a given
excitation probability $P_e$ the ratio between the heating in a cavity
and that in free space scales like the cooperativity parameter $C$.

While this diffusion can be well understood in terms of recoil kicks,
we now put the atom at a node $g=0$ and consider diffusion along the
cavity axis, $\nabla g\neq0$.
Surprisingly, although the quantum state of the cavity mode is the
vacuum $|0\rangle$ (recall that $E=0$), the atom is still
``kicked'' along the cavity axis, $|\hslash\nabla\maop|^22\kap\neq0$.
Moreover, the heating is proportional to the rate $2\kap$ at which
photons are removed from the cavity, but there are no photons to leak
out!
Another intriguing situation occurs if instead of the cavity the atom is
in the ground state $|g\rangle$. For $g\neq0$, this is achieved when in
addition to the side laser running wave a coherent field is
injected into the cavity, see Fig.~\ref{Setup}. For
$\eta=-gE/\wct$ one can show that the steady state from
(\ref{HJC}) and (\ref{Hatom-mode}) is pure
$\rho_s=|g\rangle\langle g|\!\otimes\!
|\!-\!E/\wct\rangle\langle-E/\wct|$,
where $|-E/\wct\rangle$ is a coherent state of the cavity mode.
This suppression of fluorescence occurs under a configuration which is
more general than that of Ref.~\onlinecite{Alsing92} since it accounts
for $E\neq0$, and in particular it holds true for a real lossy cavity
$\kap\neq0$.
The atom is invisible in the cavity transmission, but it is coupled.
Here, no radiation pressure acts on the atom, at rest, although the
side laser is switched on. The diffusion, however, is finite.

Such cases resemble the very intriguing situation of an
atom at a node of a standing wave, where the intensity locally vanishes
$P_e=0$, meaning that there are no fluorescence photons, but
$\DD\neq0$ \cite{Cohen92}.
The origin of the finite diffusion even when the oscillators are in
their ground state is due to the fluctuations ($\delta\sig,\delta a$) of
the atomic coherence $\sig$ and mode operator $a$, which in fact are
independent of the intensity. As noted above, they are the only
fluctuations that contribute to the diffusion in the harmonic limit, and
this is also true for any other location in space
(where the oscillators are no more in their ground state).
Following a terminology known from free space \cite{Gordon80}, one would
say that the momentum diffusion in the experimentally relevant limit
is governed by zero-point fluctuations.

We end this paper with some remarks which allow us to perceive the
physics from a broader perspective.
When one looks at the history of interpretations of the forces and their
fluctuations in free space (e.g. \cite{Cohen92} and references therein),
it is clear that the interpretations crucially depend on the spatial
modulation of the laser light. This is justified when saturation effects
are considered, because then the form of the diffusion coefficient
depends upon the structure of the laser light. But in the harmonic
limit, the diffusion reduces to the invariant form (\ref{DatTotFree}).
Adding the cavity, we found the invariant form (\ref{DatTot}), which
includes (\ref{DatTotFree}) and which is valid regardless of the way the
atom-cavity system is excited. Therefore, even if we interpreted the
particular case of radiation pressure, and communicated possible links
to the problem of the atom at the node of a standing wave, it is
appealing to interpret Eq.~(\ref{DatTot}), as much as
Eq.~(\ref{DatTotFree}) in a way that reflects such an invariance.
A general interpretation of Eq.~(\ref{DatTot}) is that of a fluctuating
atomic dipole coupled to a mean field, $|\hslash\nabla\msig|^22\gam$,
and a fluctuating cavity field coupled to a mean atomic dipole,
$|\hslash\nabla\maop|^22\kap$.
\begin{acknowledgments}
We warmly thank J. Dalibard for valuable comments.
This work was supported by the European Commission through
the Research Training Network ``CONQUEST'' and the Integrated Project
FET/QIPC ``SCALA''.
\end{acknowledgments}

\begin{thebibliography}{25}
\expandafter\ifx\csname natexlab\endcsname\relax\def\natexlab#1{#1}\fi
\expandafter\ifx\csname bibnamefont\endcsname\relax
  \def\bibnamefont#1{#1}\fi
\expandafter\ifx\csname bibfnamefont\endcsname\relax
  \def\bibfnamefont#1{#1}\fi
\expandafter\ifx\csname citenamefont\endcsname\relax
  \def\citenamefont#1{#1}\fi
\expandafter\ifx\csname url\endcsname\relax
  \def\url#1{\texttt{#1}}\fi
\expandafter\ifx\csname urlprefix\endcsname\relax\def\urlprefix{URL }\fi
\providecommand{\bibinfo}[2]{#2}
\providecommand{\eprint}[2][]{\url{#2}}

\bibitem[{\citenamefont{Mossberg et~al.}(1991)\citenamefont{Mossberg,
  Lewenstein, and Gauthier}}]{Mossberg91}
\bibinfo{author}{\bibfnamefont{T.~W.} \bibnamefont{Mossberg}},
  \bibinfo{author}{\bibfnamefont{M.}~\bibnamefont{Lewenstein}},
  \bibnamefont{and} \bibinfo{author}{\bibfnamefont{D.~J.}
  \bibnamefont{Gauthier}}, \bibinfo{journal}{\prl}
  \textbf{\bibinfo{volume}{67}}, \bibinfo{pages}{1723}
  (\bibinfo{year}{1991}).

\bibitem[{\citenamefont{Zaugg et~al.}(1994)\citenamefont{Zaugg,
Meystre, Lenz,
  and Wilkens}}]{Zaugg94}
\bibinfo{author}{\bibfnamefont{T.}~\bibnamefont{Zaugg}},
  \bibinfo{author}{\bibfnamefont{P.}~\bibnamefont{Meystre}},
  \bibinfo{author}{\bibfnamefont{G.}~\bibnamefont{Lenz}},
  \bibnamefont{and}
  \bibinfo{author}{\bibfnamefont{M.}~\bibnamefont{Wilkens}},
  \bibinfo{journal}{\pra} \textbf{\bibinfo{volume}{49}},
  \bibinfo{pages}{3011}
  (\bibinfo{year}{1994}).

\bibitem[{\citenamefont{Doherty et~al.}(1997)\citenamefont{Doherty,
Parkins,
  Tan, and Walls}}]{Doherty97}
\bibinfo{author}{\bibfnamefont{A.~C.} \bibnamefont{Doherty}},
  \bibinfo{author}{\bibfnamefont{A.~S.} \bibnamefont{Parkins}},
  \bibinfo{author}{\bibfnamefont{S.~M.} \bibnamefont{Tan}},
  \bibnamefont{and}
  \bibinfo{author}{\bibfnamefont{D.~F.} \bibnamefont{Walls}},
  \bibinfo{journal}{\pra} \textbf{\bibinfo{volume}{56}},
  \bibinfo{pages}{833}
  (\bibinfo{year}{1997}).

\bibitem[{\citenamefont{Horak et~al.}(1997)\citenamefont{Horak,
Hechenblaikner,
  Gheri, Stecher, and Ritsch}}]{Horak97}
\bibinfo{author}{\bibfnamefont{P.}~\bibnamefont{Horak}},
  \bibinfo{author}{\bibfnamefont{G.}~\bibnamefont{Hechenblaikner}},
  \bibinfo{author}{\bibfnamefont{K.~M.} \bibnamefont{Gheri}},
  \bibinfo{author}{\bibfnamefont{H.}~\bibnamefont{Stecher}},
  \bibnamefont{and}
  \bibinfo{author}{\bibfnamefont{H.}~\bibnamefont{Ritsch}},
  \bibinfo{journal}{\prl} \textbf{\bibinfo{volume}{79}},
  \bibinfo{pages}{4974}
  (\bibinfo{year}{1997}).

\bibitem[{\citenamefont{Hechenblaikner
  et~al.}(1998)\citenamefont{Hechenblaikner, Gangl, Horak, and
  Ritsch}}]{Hechenblaikner98}
\bibinfo{author}{\bibfnamefont{G.}~\bibnamefont{Hechenblaikner}},
  \bibinfo{author}{\bibfnamefont{M.}~\bibnamefont{Gangl}},
  \bibinfo{author}{\bibfnamefont{P.}~\bibnamefont{Horak}},
  \bibnamefont{and}
  \bibinfo{author}{\bibfnamefont{H.}~\bibnamefont{Ritsch}},
  \bibinfo{journal}{\pra} \textbf{\bibinfo{volume}{58}},
  \bibinfo{pages}{3030}
  (\bibinfo{year}{1998}).

\bibitem[{\citenamefont{{Vuleti{\'c}} and {Chu}}(2000)}]{Vuletic00}
\bibinfo{author}{\bibfnamefont{V.}~\bibnamefont{{Vuleti{\'c}}}}
  \bibnamefont{and}
  \bibinfo{author}{\bibfnamefont{S.}~\bibnamefont{{Chu}}},
  \bibinfo{journal}{Phys. Rev. Lett.}
  \textbf{\bibinfo{volume}{84}},
  \bibinfo{pages}{3787} (\bibinfo{year}{2000}).

\bibitem[{\citenamefont{van Enk et~al.}(2001)\citenamefont{van Enk,
McKeever,
  Kimble, and Ye}}]{Enk01}
\bibinfo{author}{\bibfnamefont{S.~J.} \bibnamefont{van Enk}},
  \bibinfo{author}{\bibfnamefont{J.}~\bibnamefont{McKeever}},
  \bibinfo{author}{\bibfnamefont{H.~J.} \bibnamefont{Kimble}},
  \bibnamefont{and}
  \bibinfo{author}{\bibfnamefont{J.}~\bibnamefont{Ye}},
  \bibinfo{journal}{\pra} \textbf{\bibinfo{volume}{64}},
  \bibinfo{pages}{013407} (\bibinfo{year}{2001}).

\bibitem[{\citenamefont{{Vuleti{\'c}}
et~al.}(2001)\citenamefont{{Vuleti{\'c}},
  {Chan}, and {Black}}}]{Vuletic01}
\bibinfo{author}{\bibfnamefont{V.}~\bibnamefont{{Vuleti{\'c}}}},
  \bibinfo{author}{\bibfnamefont{H.~W.} \bibnamefont{{Chan}}},
  \bibnamefont{and} \bibinfo{author}{\bibfnamefont{A.~T.}
  \bibnamefont{{Black}}}, \bibinfo{journal}{\pra}
  \textbf{\bibinfo{volume}{64}}, \bibinfo{pages}{033405}
  (\bibinfo{year}{2001}).

\bibitem[{\citenamefont{Domokos et~al.}(2001)\citenamefont{Domokos,
Horak, and
  Ritsch}}]{Domokos01}
\bibinfo{author}{\bibfnamefont{P.}~\bibnamefont{Domokos}},
  \bibinfo{author}{\bibfnamefont{P.}~\bibnamefont{Horak}},
  \bibnamefont{and}
  \bibinfo{author}{\bibfnamefont{H.}~\bibnamefont{Ritsch}},
  \bibinfo{journal}{J. Phys B} \textbf{\bibinfo{volume}{34}},
  \bibinfo{pages}{187} (\bibinfo{year}{2001}).

\bibitem[{\citenamefont{Fischer et~al.}(2001)\citenamefont{Fischer,
Maunz,
  Puppe, Pinkse, and Rempe}}]{Fischer01}
\bibinfo{author}{\bibfnamefont{T.}~\bibnamefont{Fischer}},
  \bibinfo{author}{\bibfnamefont{P.}~\bibnamefont{Maunz}},
  \bibinfo{author}{\bibfnamefont{T.}~\bibnamefont{Puppe}},
  \bibinfo{author}{\bibfnamefont{P.~W.~H.} \bibnamefont{Pinkse}},
  \bibnamefont{and}
  \bibinfo{author}{\bibfnamefont{G.}~\bibnamefont{Rempe}},
  \bibinfo{journal}{New J. Phys.} \textbf{\bibinfo{volume}{3}},
  \bibinfo{pages}{11.1} (\bibinfo{year}{2001}).

\bibitem[{\citenamefont{Domokos et~al.}(2002)\citenamefont{Domokos,
Salzburger,
  and Ritsch}}]{Domokos02}
\bibinfo{author}{\bibfnamefont{P.}~\bibnamefont{Domokos}},
  \bibinfo{author}{\bibfnamefont{T.}~\bibnamefont{Salzburger}},
  \bibnamefont{and}
  \bibinfo{author}{\bibfnamefont{H.}~\bibnamefont{Ritsch}},
  \bibinfo{journal}{\pra} \textbf{\bibinfo{volume}{66}},
  \bibinfo{pages}{043406} (\bibinfo{year}{2002}).

\bibitem[{\citenamefont{Murr}(2003)}]{Murr03}
\bibinfo{author}{\bibfnamefont{K.}~\bibnamefont{Murr}},
\bibinfo{journal}{J.
  Phys. B} \textbf{\bibinfo{volume}{36}}, \bibinfo{pages}{2515}
  (\bibinfo{year}{2003}).

\bibitem[{\citenamefont{Domokos
et~al.}(2004)\citenamefont{Domokos, Vukics, and
  Ritsch}}]{Domokos04}
\bibinfo{author}{\bibfnamefont{P.}~\bibnamefont{Domokos}},
  \bibinfo{author}{\bibfnamefont{A.}~\bibnamefont{Vukics}},
  \bibnamefont{and}
  \bibinfo{author}{\bibfnamefont{H.}~\bibnamefont{Ritsch}},
  \bibinfo{journal}{\prl} \textbf{\bibinfo{volume}{92}},
  \bibinfo{pages}{103601} (\bibinfo{year}{2004}).

\bibitem[{\citenamefont{Maunz et~al.}(2004)\citenamefont{Maunz, Puppe,
  Schuster, Syassen, Pinkse, and Rempe}}]{Maunz04}
\bibinfo{author}{\bibfnamefont{P.}~\bibnamefont{Maunz}},
  \bibinfo{author}{\bibfnamefont{T.}~\bibnamefont{Puppe}},
  \bibinfo{author}{\bibfnamefont{I.}~\bibnamefont{Schuster}},
  \bibinfo{author}{\bibfnamefont{N.}~\bibnamefont{Syassen}},
  \bibinfo{author}{\bibfnamefont{P.~W.~H.} \bibnamefont{Pinkse}},
  \bibnamefont{and}
  \bibinfo{author}{\bibfnamefont{G.}~\bibnamefont{Rempe}},
  \bibinfo{journal}{Nature (London)}
  \textbf{\bibinfo{volume}{428}},
  \bibinfo{pages}{50} (\bibinfo{year}{2004}).

\bibitem[{\citenamefont{{Nu{\ss}mann}
et~al.}(2005)\citenamefont{{Nu{\ss}mann},
  {Murr}, {Hijlkema}, {Weber}, {Kuhn}, and {Rempe}}}]{Nussmann05}
\bibinfo{author}{\bibfnamefont{S.}~\bibnamefont{{Nu{\ss}mann}}},
  \bibinfo{author}{\bibfnamefont{K.}~\bibnamefont{{Murr}}},
  \bibinfo{author}{\bibfnamefont{M.}~\bibnamefont{{Hijlkema}}},
  \bibinfo{author}{\bibfnamefont{B.}~\bibnamefont{{Weber}}},
  \bibinfo{author}{\bibfnamefont{A.}~\bibnamefont{{Kuhn}}},
  \bibnamefont{and}
  \bibinfo{author}{\bibfnamefont{G.}~\bibnamefont{{Rempe}}},
  \bibinfo{journal}{Nature Phys.} \textbf{\bibinfo{volume}{1}},
  \bibinfo{pages}{122} (\bibinfo{year}{2005}).

\bibitem[{\citenamefont{M{\"u}nstermann
  et~al.}(1999)\citenamefont{M{\"u}nstermann, Fischer, Maunz, Pinkse,
  and
  Rempe}}]{muenstermann99}
\bibinfo{author}{\bibfnamefont{P.}~\bibnamefont{M{\"u}nstermann}},
  \bibinfo{author}{\bibfnamefont{T.}~\bibnamefont{Fischer}},
  \bibinfo{author}{\bibfnamefont{P.}~\bibnamefont{Maunz}},
  \bibinfo{author}{\bibfnamefont{P.~W.~H.} \bibnamefont{Pinkse}},
  \bibnamefont{and}
  \bibinfo{author}{\bibfnamefont{G.}~\bibnamefont{Rempe}},
  \bibinfo{journal}{\prl} \textbf{\bibinfo{volume}{82}},
  \bibinfo{pages}{3791}
  (\bibinfo{year}{1999}).

\bibitem[{\citenamefont{Ye et~al.}(1999)\citenamefont{Ye, Vernooy, and
  Kimble}}]{Ye99}
\bibinfo{author}{\bibfnamefont{J.}~\bibnamefont{Ye}},
  \bibinfo{author}{\bibfnamefont{D.~W.} \bibnamefont{Vernooy}},
  \bibnamefont{and} \bibinfo{author}{\bibfnamefont{H.~J.}
  \bibnamefont{Kimble}}, \bibinfo{journal}{\prl}
  \textbf{\bibinfo{volume}{83}},
  \bibinfo{pages}{4987} (\bibinfo{year}{1999}).

\bibitem[{\citenamefont{McKeever et~al.}(2003)\citenamefont{McKeever,
Buck,
  Boozer, Kuzmich, N{\"a}gerl, Stamper-Kurn, and Kimble}}]{McKeever03}
\bibinfo{author}{\bibfnamefont{J.}~\bibnamefont{McKeever}},
  \bibinfo{author}{\bibfnamefont{J.~R.} \bibnamefont{Buck}},
  \bibinfo{author}{\bibfnamefont{A.~D.} \bibnamefont{Boozer}},
  \bibinfo{author}{\bibfnamefont{A.}~\bibnamefont{Kuzmich}},
  \bibinfo{author}{\bibfnamefont{H.-C.} \bibnamefont{N{\"a}gerl}},
  \bibinfo{author}{\bibfnamefont{D.~M.} \bibnamefont{Stamper-Kurn}},
  \bibnamefont{and} \bibinfo{author}{\bibfnamefont{H.~J.}
  \bibnamefont{Kimble}}, \bibinfo{journal}{\prl}
  \textbf{\bibinfo{volume}{90}},
  \bibinfo{pages}{133602} (\bibinfo{year}{2003}).

\bibitem[{\citenamefont{Boca et~al.}(2004)\citenamefont{Boca, Miller,
Birnbaum,
  Boozer, McKeever, and Kimble}}]{Boca04}
\bibinfo{author}{\bibfnamefont{A.}~\bibnamefont{Boca}},
  \bibinfo{author}{\bibfnamefont{R.}~\bibnamefont{Miller}},
  \bibinfo{author}{\bibfnamefont{K.~M.} \bibnamefont{Birnbaum}},
  \bibinfo{author}{\bibfnamefont{A.~D.} \bibnamefont{Boozer}},
  \bibinfo{author}{\bibfnamefont{J.}~\bibnamefont{McKeever}},
  \bibnamefont{and}
  \bibinfo{author}{\bibfnamefont{H.~J.} \bibnamefont{Kimble}},
  \bibinfo{journal}{\prl} \textbf{\bibinfo{volume}{93}},
  \bibinfo{pages}{233603} (\bibinfo{year}{2004}).

\bibitem[{\citenamefont{Maunz et~al.}(2005)\citenamefont{Maunz, Puppe,
  Schuster, Syassen, Pinkse, and Rempe}}]{Maunz05}
\bibinfo{author}{\bibfnamefont{P.}~\bibnamefont{Maunz}},
  \bibinfo{author}{\bibfnamefont{T.}~\bibnamefont{Puppe}},
  \bibinfo{author}{\bibfnamefont{I.}~\bibnamefont{Schuster}},
  \bibinfo{author}{\bibfnamefont{N.}~\bibnamefont{Syassen}},
  \bibinfo{author}{\bibfnamefont{P.~W.~H.} \bibnamefont{Pinkse}},
  \bibnamefont{and}
  \bibinfo{author}{\bibfnamefont{G.}~\bibnamefont{Rempe}},
  \bibinfo{journal}{\prl} \textbf{\bibinfo{volume}{94}},
  \bibinfo{pages}{033002} (\bibinfo{year}{2005}).

\bibitem[{\citenamefont{Cook}(1980)}]{Cook80}
\bibinfo{author}{\bibfnamefont{R.~J.} \bibnamefont{Cook}},
  \bibinfo{journal}{\pra} \textbf{\bibinfo{volume}{22}},
  \bibinfo{pages}{1078}
  (\bibinfo{year}{1980}).

\bibitem[{\citenamefont{Cohen-Tannoudji}(1992)}]{Cohen92}
\bibinfo{author}{\bibfnamefont{C.}~\bibnamefont{Cohen-Tannoudji}}, in
  \emph{\bibinfo{booktitle}{Fundamental Systems in Quantum Optics,
  Les Houches,
  Session LIII, 1990}}, edited by
  \bibinfo{editor}{\bibfnamefont{J.}~\bibnamefont{Dalibard}},
  \bibinfo{editor}{\bibfnamefont{J.~M.} \bibnamefont{Raimond}},
  \bibnamefont{and}
  \bibinfo{editor}{\bibfnamefont{J.}~\bibnamefont{Zinn-Justin}}
  (\bibinfo{publisher}{Elsevier Science},
  \bibinfo{address}{North-Holland,
  Amsterdam}, \bibinfo{year}{1992}), p.~\bibinfo{pages}{1}.

\bibitem[{\citenamefont{Gordon and Ashkin}(1980)}]{Gordon80}
\bibinfo{author}{\bibfnamefont{J.~P.} \bibnamefont{Gordon}}
\bibnamefont{and}
  \bibinfo{author}{\bibfnamefont{A.}~\bibnamefont{Ashkin}},
  \bibinfo{journal}{\pra} \textbf{\bibinfo{volume}{21}},
  \bibinfo{pages}{1606}
  (\bibinfo{year}{1980}).

\bibitem[{\citenamefont{Dalibard
and Cohen-Tannoudji}(1985)}]{Dalibard85}
\bibinfo{author}{\bibfnamefont{J.}~\bibnamefont{Dalibard}}
\bibnamefont{and}
  \bibinfo{author}{\bibfnamefont{C.}~\bibnamefont{Cohen-Tannoudji}},
  \bibinfo{journal}{\josab} \textbf{\bibinfo{volume}{2}},
  \bibinfo{pages}{1707}
  (\bibinfo{year}{1985}).

\bibitem[{\citenamefont{Alsing et~al.}(1992)\citenamefont{Alsing,
Cardimona,
  and Carmichael}}]{Alsing92}
\bibinfo{author}{\bibfnamefont{P.~M.} \bibnamefont{Alsing}},
  \bibinfo{author}{\bibfnamefont{D.~A.} \bibnamefont{Cardimona}},
  \bibnamefont{and} \bibinfo{author}{\bibfnamefont{H.~J.}
  \bibnamefont{Carmichael}}, \bibinfo{journal}{\pra}
  \textbf{\bibinfo{volume}{45}}, \bibinfo{pages}{1793}
  (\bibinfo{year}{1992}).

\end{thebibliography}

\end{document}